\documentclass[prl,twocolumn,showpacs]{revtex4}

\usepackage{bm}
\usepackage{graphicx}

\newcommand*{\be}{\begin{equation}}
\newcommand*{\ee}{\end{equation}}

\begin{document}

\title{Correlated defects, metal-insulator transition, and magnetic order\\
in ferromagnetic semiconductors}

\author{C. Timm}
\email{timm@physik.fu-berlin.de}
\author{F. Sch\"afer}
\author{F. von Oppen}
\affiliation{Institut f\"ur Theoretische Physik, Freie Universit\"at Berlin,
Arnimallee 14, D-14195 Berlin, Germany}

\date{January 22, 2002}

\begin{abstract}
The effect of disorder on transport and magnetization in ferromagnetic
III--V semiconductors, in particular (Ga,Mn)As, is studied theoretically.
We show that Coulomb-induced correlations of the defect positions are
crucial for the transport and magnetic properties of these highly
compensated materials. We employ Monte Carlo simulations to obtain the
correlated defect distributions. Exact diagonalization gives reasonable
results for the spectrum of valence-band holes and the metal-insulator
transition only for correlated disorder. Finally, we show that the
mean-field magnetization also depends crucially on defect correlations.
\end{abstract}

\pacs{75.50.Pp, 71.30.+h, 72.20.Ee}

\maketitle



\emph{Introduction}.---Recently, there has been substantial interest in
diluted ferromagnetic III--V semiconductors due to observations of
relatively high Curie temperatures \cite{Ohno,Ohno1,Reed}. This makes these
materials promising for applications as well as interesting from the
physics point of view. They could allow the incorporation of ferromagnetic
elements into semiconductor devices, and thus the integration of processing
and magnetic storage on a single chip. To be specific, we consider here
manganese-doped GaAs. The properties of this material rely on the dual role
played by the Mn impurities: They carry a local spin due to their
half-filled d-shell and dope the system with holes, which mediate a
ferromagnetic indirect exchange interaction between the spins. Furthermore,
the materials are highly compensated presumably due to antisite defects (As
substituted for Ga) \cite{Potashnik}, which drives them towards a
metal-insulator transition (MIT). It is clearly important to understand the
interplay between transport properties, magnetic ordering, and the defect
configuration. In this letter we show that \emph{correlated} defects are
required for a description consistent with experiments.

%


There is a growing body of theoretical work on (Ga,Mn)As \cite{Koenigrev}.
Most approaches either start from the heavily doped regime
\cite{Dietl97,Jungwirth,Koenig,Abolfath,Dietl,CDM,Schlie} or from the
weak-doping limit \cite{BB}. In the case of heavy doping
\cite{Dietl97,Jungwirth,Koenig,Abolfath,Dietl,CDM,Schlie} one considers
valence-band holes moving in the disorder potential of the defects. For
large hole concentration the Fermi energy $E_F$ lies deep in the valence
band compared to the Coulomb-potential fluctuations and the latter are
neglected. On the other hand, for light doping \cite{BB} the local
impurity states overlap only weakly, forming an impurity band. This model
is problematic since for the typical doping range the impurity band would
be much broader than the energy gap \cite{comment}. Both approaches have in
common that the effect of antisites on the disorder and correlations of
defects are neglected.


The outline of the remainder of this paper is as follows. First, we discuss
the configuration of Mn impurities and antisite defects with the help of
Monte Carlo simulations and obtain the disorder potential and its spatial
correlations. Then, we derive the spectrum and localization properties of
holes in this potential and discuss the MIT. Finally, we calculate the
polarizations of Mn and hole spins within a selfconsistent mean-field
theory.



\emph{Defect configurations}.---In the theory of doped semiconductors one
often assumes the dopants to be randomly distributed \cite{SE,Koenigrev}.
However, the present system is highly compensated. This leads to defects of
either charge being present (relative to Ga, Mn impurities have charge $-e$
while antisites carry $+2e$) and to a low density of carriers, which only
weakly screen the Coulomb interaction between the charged defects
\cite{SE}. In the resulting \emph{Coulomb plasma} the defect positions are
expected to be highly correlated. We study here the limiting case where the
defects come into thermal equilibrium during growth \cite{Keldysh}. This is
motivated by experiments \cite{Potashnik} suggesting that defects diffuse
rapidly at $250^\circ$C, which is a typical growth and annealing
temperature for (Ga,Mn)As grown by molecular beam epitaxy. The resulting
configuration is assumed to be quenched at low temperatures.



To find typical defect configurations close to equilibrium we perform Monte
Carlo simulations for the classical Coulomb system of Mn impurities and
antisites on the Ga sublattice with the Hamiltonian $H=1/2\, \sum_{i,j}
q_iq_j/(\epsilon r_{ij})\, e^{-r_{ij}/r_{\mathrm{scr}}}$, where $q_i$ are
the defect charges, $r_{ij}$ is their separation, and $\epsilon\cong 11$.
The screening length $r_{\mathrm{scr}}$ is obtained from nonlinear
screening theory \cite{SE}. $r_{\mathrm{scr}}$ is much larger than the
nearest-neighbor separation of Ga sites for realistic parameters so that it
hardly affects the small-scale defect correlations relevant here. We employ
the Metropolis algorithm at $250^\circ$C for systems of
$20\times20\times20$ conventional cubic unit cells with periodic boundary
conditions, unless stated otherwise.



\begin{figure}[ht]
\includegraphics[width=3.20in]{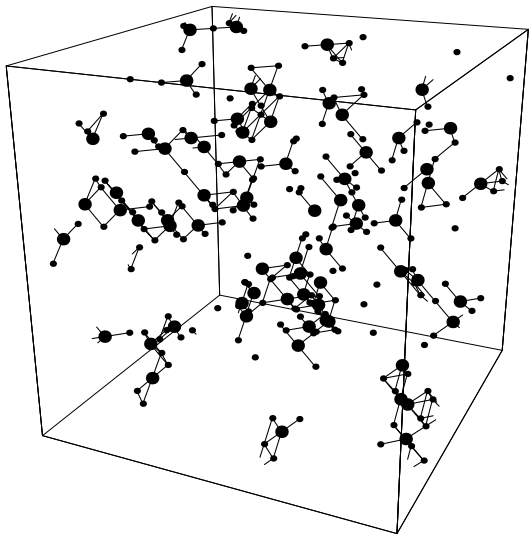}
\caption{\label{fig.conf}Typical equilibrium configuration at $250^\circ$C
for 5\% Mn doping and 0.3 holes per Mn. Only the defects are shown, large
(small) circles denote antisite (Mn) defects. Defects at nearest neighbor
Ga sites are connected by a bond. For clarity we show only
$10\times10\times10$ conventional unit cells.}
\end{figure}

\begin{figure}[ht]
\includegraphics[width=3.20in]{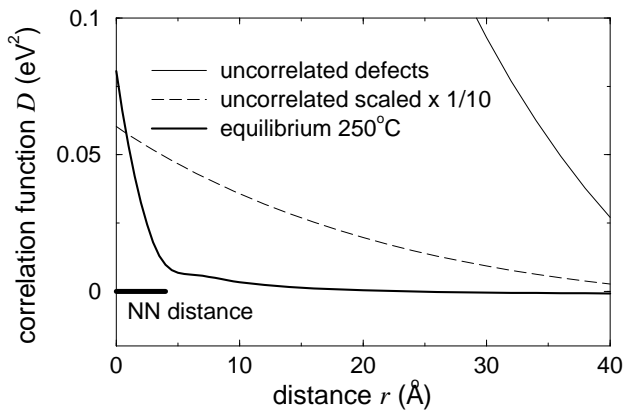}
\caption{\label{fig.correl}Potential correlation function $D(r)$ for 5\% Mn
and 0.3 holes per Mn, both for uncorrelated defects and for equilibrium at
the growth temperature. The nearest-neighbor Ga site separation of
3.99\,\AA{} is also indicated.}
\end{figure}

A typical result for Mn concentration $x=0.05$ and $p=0.3$ holes per Mn, as
suggested by experiments \cite{Ohno}, is shown in Fig.~\ref{fig.conf}. The
concentration of antisites is obtained from $x$ and $p$ under the
assumption of charge neutrality. Most of the defects have formed clusters.
This leads to screening of the disorder potential by the defects, as
can be seen from the potential correlation function $D(r)\equiv\langle
V({\bf r})\,V({\bf r}') \rangle_{\scriptscriptstyle |{\bf r}-{\bf r}'|=r} -
\langle V \rangle^2$ plotted in Fig.~\ref{fig.correl} for $x=0.05$ and
$p=0.3$. Note that $\Delta V\equiv\sqrt{D(0)}$ is the width of the
distribution of $V(\mathbf{r})$. For correlated defects, $D(r)$ and thus
$\Delta V$ are strongly reduced and the Coulomb potential is screened on
the much shorter length scale $r_{\mathrm{ion}}$, which is of the order of
the nearest-neighbor Ga site separation \cite{rem.rscr}.
(This ionic
screening is only possible due to the presence of defects of either charge.
Thus it does not take place, \emph{e.g.}, in ferromagnetic II--VI
semiconductors since Mn is isovalent in these materials.)
Importantly, however, $\Delta V\approx 0.284\,$eV is still \emph{not} small
compared to the Fermi energy $|E_F|\approx 0.329\,$eV.

Recent high resolution x-ray diffraction \cite{Schott} as well as
resistivity and magnetization measurements \cite{Potashnik} support the
formation of small clusters. The more direct determination of the local
structure around Mn dopants by extended x-ray-absorption fine structure
\cite{Shioda} does not give direct information on the abundance of specific
elements around the Mn. However, the observed loss of order around Mn is
certainly consistent with cluster formation. Finally, we show in the
following that experimental results for the band gap, transport, and the
magnetization \cite{Ohno,Beschoten} strongly support the clustering.



\emph{Hole states}.---To find the spectrum and localization properties of
valence-band holes moving in the disorder potential $V(\mathbf{r})$, we
start from the Hamiltonian
$H=-\sum_i(\hbar^2/2m^\ast)\,\nabla_i^2+V(\mathbf{r}_i)$, where we use the
envelope function and parabolic-band approximations \cite{rem.KL}. The
calculations are done for spin-less holes, which is justified since the
additional disorder introduced by the exchange interaction with the Mn
spins is much smaller than $\Delta V$.



\begin{figure}
\includegraphics[width=3.20in]{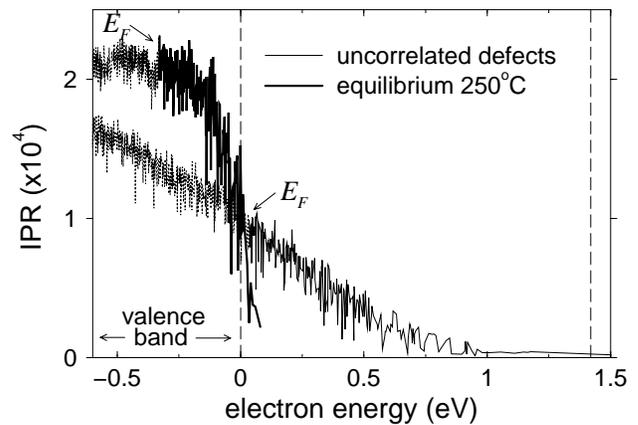}
\caption{\label{fig.ipr.correl}IPR as a function of energy at the
valence-band edge for 5\% Mn and 0.3 holes per Mn. The thin (heavy) solid
line denotes empty states for uncorrelated (correlated) defects, the
dotted lines denote occupied states.}
\end{figure}

The Hamiltonian is solved by exact diagonalization in a plane-wave basis of
$13^3=2197$ states \cite{rem.number}, giving the energy spectrum and
normalized eigenfunctions. The Fermi energy $E_F$ is obtained from the hole
concentration $px$. We also obtain the inverse participation ratio
$\mathrm{IPR}(n)=1/\sum_{\bf r} |\psi_n({\bf r})|^4$ of the states
$\psi_n({\bf r})$. The IPR allows to estimate the position of the mobility
edge, since it is of the order of the volume for extended states but is
essentially independent of system size for localized states.
Figure \ref{fig.ipr.correl} shows a comparison of the IPR as a function of
energy for uncorrelated defects and for the equilibrium
configuration obtained above, for 5\% Mn and $0.3$ holes per Mn.
The valence-band edge is
strongly smeared out by uncorrelated defects so that the energy gap is
completely filled, which is not observed experimentally, whereas correlated
defects only lead to a small tail in the gap. We have also studied the
dependence of the IPR on system size $L^3$ over the limited feasible range
in $L$, finding that $\mathrm{IPR}(n)\propto L^3$ for the flat part of the
curve, signifying extended states, whereas the IPR is independent of $L$
for states sufficiently far in the tail, which are thus localized. Hence,
the mobility edge lies on the slope of the curve. The states are much more
localized for uncorrelated defects, due to the larger disorder potential,
whereas for correlated defects the states close to $E_F$ are clearly
extended. Thus the spectrum and the localization properties agree
with experiments \cite{Ohno} only for correlated defects.


\begin{figure}
\includegraphics[width=3.20in]{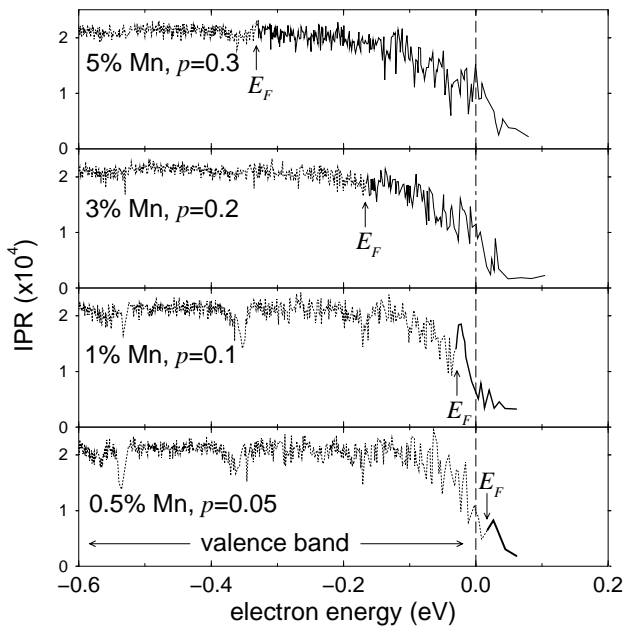}
\caption{\label{fig.ipr.dope}IPR as a function of energy at
the valence-band edge for various Mn dopings and hole concentrations $p$ per
Mn taken from experiment \cite{Ohno,Matsukura}.}
\end{figure}

In Fig.~\ref{fig.ipr.dope} we show the IPR as a function of energy for
correlated defects at various Mn dopings and hole concentrations. The
hole concentration for given Mn doping was chosen in
accordance with experiments \cite{Ohno,Matsukura}. The spectrum and the IPR
change little with Mn doping, since the potential screened by
the defects on the length scale $r_{\mathrm{ion}}$ is much less affected
by a change of concentration than the bare Coulomb potential.



The dominant effect comes from the hole concentration. Our result for the
scaling of the IPR suggests that for 5\% and 3\% Mn the states at $E_F$
are extended and we expect metallic behavior. For 1\% Mn $E_F$ is
close to the mobility edge, while for 0.5\% Mn the states at
$E_F$ are localized. From this we estimate the MIT to take place at a Mn
concentration of the order of 1\%. This result is in reasonable agreement with
experiments \cite{Ohno}. For comparison, uncorrelated defects would lead to
an MIT at about 5\% Mn, see Fig.~\ref{fig.ipr.correl}.



\emph{Magnetization}.---Finally, we discuss the spontaneous magnetization
and its dependence of the defect configuration. The exchange
interaction between hole and Mn spins is described by the Hamiltonian
$H = -\sum_i(\hbar^2/2m^\ast)\,\nabla_i^2 +
V(\mathbf{r}_i) - J_{\mathrm{pd}} \sum_{i,l} \mathbf{s}_i \cdot
\mathbf{S}_l\,\delta(\mathbf{r}_i-\mathbf{R}_l)$,
where $V(\mathbf{r})$ is the disorder potential obtained above,
$J_{\mathrm{pd}}\approx -45\,$eV\,\AA$^3$
is the exchange integral \cite{Okaba,Bhatta,Abolfath}, $\mathbf{s}_i$
and $\mathbf{S}_l$ are the hole and Mn spin operators, respectively, and
$\mathbf{R}_l$ are the Mn impurity positions. In second quantized form,
the Hamiltonian reads
\begin{eqnarray*}
H & = & \sum_{n\sigma} c_{n\sigma}^\dagger\,\epsilon_n\,c_{n\sigma} \\
& & {\!\!\!}- J_{\mathrm{pd}}
    \sum_{n\sigma,n'\sigma'} \sum_l c_{n\sigma}^\dagger\,
    \psi_n^\ast(\mathbf{R}_l)\,
    \frac{\bm{\tau}_{\sigma\sigma'}}{2} \cdot \mathbf{S}_l\,
    \psi_{n'}(\mathbf{R}_l)\, c_{n'\sigma'} ,
\end{eqnarray*}
where $\epsilon_n$ are the eigenenergies obtained in the absence of
exchange as discussed above, $\psi_n(\mathbf{r})$ are the corresponding
eigenfunctions, and $\bm{\tau}$ is the vector of Pauli matrices. The
exchange interaction is decoupled at the mean-field level, introducing the
averaged Mn spin polarizations
$\mathbf{M}_l\equiv\langle\mathbf{S}_l\rangle$ and hole spin polarizations
at the Mn sites, $\mathbf{m}_l\equiv\langle\sum_{n\sigma,n'\sigma'}
c_{n\sigma}^\dagger\psi_n^\ast(\mathbf{R}_l)
(\bm{\tau}_{\sigma\sigma'}/2)\psi_{n'}(\mathbf{R}_l)c_{n'\sigma'}\rangle$.
We do \emph{not} perform a spatial average, \textit{i.e.}, the disorder is
retained. The hole and Mn sectors can now be diagonalized separately,
$\mathbf{M}_l$ and $\mathbf{m}_l$ are calculated, and the procedure is
iterated. This selfconsistent mean-field theory is similar to the one
employed by Berciu and Bhatt \cite{BB}. The main difference is that we
start from realistic hole states for the disorder potential.



\begin{figure}
\includegraphics[width=3.20in]{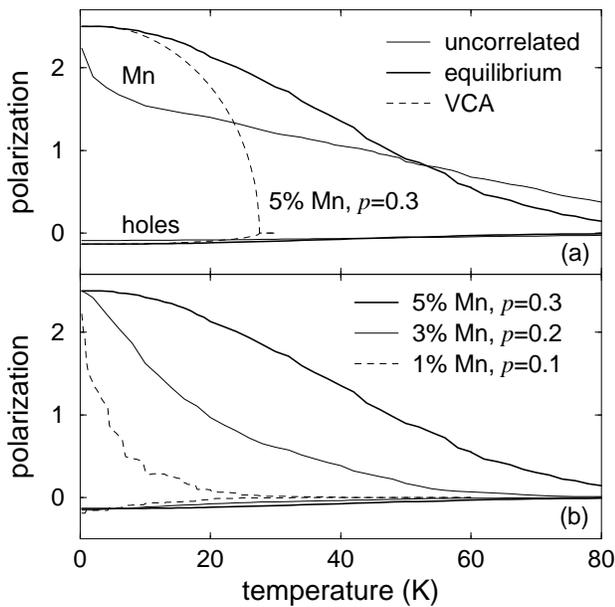}
\caption{\label{fig.magn}(a) Magnetization as a function of temperature for
5\% Mn and 0.3 holes per Mn, both for uncorrelated and correlated defects.
The results denoted by ``VCA'' neglect disorder. (b) Magnetization for
various Mn and hole concentrations (with correlated defects).}
\end{figure}

Figure \ref{fig.magn}(a) shows the Mn and hole spin polarizations for
$x=0.05$ and $p=0.3$. Also shown are the polarizations obtained neglecting
disorder by using plane waves for the hole states and the virtual crystal
approximation (VCA) for the Mn spins \cite{Jungwirth,Koenigrev,rem.KL}. We
see that disorder strongly enhances $T_c$ \cite{BB}. This increase is
easily understood: $T_c$ is determined by the indirect exchange interaction
of nearest-neighbor Mn pairs, which is inversely proportional to the
separation \cite{Dietl97,Jungwirth,Koenig}. Without disorder, the typical
distance is $r_{\mathrm{vca}}\sim n_{\mathrm{Mn}}^{-1/3}$, where
$n_{\mathrm{Mn}}$ is the density of Mn impurites. For $x=0.05$,
$r_{\mathrm{vca}}\sim 9.67\,$\AA. On the other hand, with disorder the
nearest-neighbor separation on the Ga sublattice is
$r_{\mathrm{dis}}=3.99\,$\AA. Therefore, the indirect exchange, and thus
$T_c$, is larger by a factor of order
$r_{\mathrm{vca}}/r_{\mathrm{dis}}\sim 2.4$ in the disordered case.

For \emph{correlated} defects the polarization curve looks more
``mean-field-like,'' in qualitative agreement with experiments
\cite{Ohno,Dietl,Beschoten}. This can be explained by the extended nature
of the relevant hole states for correlated defects, which leads to a
long-range indirect exchange interaction. Thus the effective field seen by
a Mn spin is averaged over many other Mn spins and the spatial fluctuations
of this effective field are small. On the other hand, for
\emph{uncorrelated} defects the holes are partly localized, the indirect
exchange is of shorter range, and the fluctuations are larger. We believe
that this leads to the smeared-out magnetization curve.
It is clearly important to study the effect of fluctuations on
the magnetization. We expect fluctuations to reduce $T_c$ in particular for
the uncorrelated case.
Figure \ref{fig.magn}(b) shows that $T_c$ is reduced for smaller Mn and
hole concentrations and that the typical temperature scale is
approximately proportional to $x$, in accordance with
experiments \cite{Ohno}. Note also that in all cases the holes are only
partially polarized.



To conclude, we have obtained typical defect configurations in (Ga,Mn)As
and studied the effect of the resulting disorder potential on the
valence-band holes. The strong Coulomb interactions of charged defects
leads to the formation of defect clusters during growth. The previously
neglected antisite defects are crucial for this effect. We have shown that
such a correlated defect distribution is required to understand the hole
spectrum, the metal-insulator transition, and the shape of the
magnetization curves. Our results should also help to clarify why the
properties of ferromagnetic semiconductors depend so strongly on details of
the growth process.---We profited from discussions with P. J. Jensen, J.
K\"onig, J. Schliemann, and in particular M. E. Raikh, who emphasized to us
the importance of compensation.

\emph{Note added}.---Recently, Yang and MacDonald \cite{Yang} have studied
the effect of the hole-hole interaction, assuming a random distribution of
Mn dopants and antisites.


\begin{thebibliography}{99}

\bibitem{Ohno}H. Ohno, Science \textbf{281}, 951 (1998);
H. Ohno, J. Magn.\ Magn.\ Mat.\ \textbf{200}, 110 (1999);
H. Ohno and F. Matsukura, Solid State Commun.\ \textbf{117}, 179 (2001).

\bibitem{Ohno1}H. Ohno \textit{et al.}, Appl.\ Phys.\ Lett.\ \textbf{69},
363 (1996); F. Matsukura \textit{et al.}, Phys.\ Rev.\ B \textbf{57}, R2037
(1998).

\bibitem{Reed}M. L. Reed \textit{et al.}, Appl.\ Phys.\ Lett.\ \textbf{79},
3473 (2001).



\bibitem{Potashnik}S. J. Potashnik \textit{et al.}, Appl.\ Phys.\ Lett.\
\textbf{79}, 1495 (2001); see also R. C. Lutz \textit{et al.}, Physica B
\textbf{273--274}, 722 (1999).

\bibitem{Koenigrev}For a review see J. K\"onig \textit{et al.}, to be
published in \textit{Electronic Structure and Magnetism of Complex
Materials}, edited by D. J. Singh and D. A. Papaconstantopoulos (Springer,
Berlin, 2002), \eprint{cond-mat/0111314}.

\bibitem{Dietl97}T. Dietl, A. Haury, and Y. Merle d'Aubign\'e, Phys.\ Rev.\
B \textbf{55}, R3347 (1997); T. Dietl \textit{et al.}, Science
\textbf{287}, 1019 (2000).

\bibitem{Jungwirth}T. Jungwirth \textit{et al.}, Phys.\ Rev.\ B
\textbf{59}, 9818 (1999).

\bibitem{Koenig}J. K\"onig, H.-H. Lin, and A. H. MacDonald, Phys.\ Rev.\
Lett.\ \textbf{84}, 5628 (2000); J. K\"onig, T. Jungwirth, and A. H.
MacDonald, Phys.\ Rev.\ B \textbf{64}, 184423 (2001).

\bibitem{Abolfath}M. Abolfath \textit{et al.}, Phys.\ Rev.\ B \textbf{63},
054418 (2001).

\bibitem{Dietl}T. Dietl, H. Ohno, and F. Matsukura, Phys.\ Rev.\ B
\textbf{63}, 195205 (2001).

\bibitem{CDM}A. Chattopadhyay, S. Das Sarma, and A. J. Millis,
Phys.\ Rev.\ Lett.\ \textbf{87}, 227202 (2001).

\bibitem{Schlie}J. Schliemann and A. H. MacDonald,
Phys.\ Rev.\ Lett.\ \textbf{88}, 137201 (2002).


\bibitem{BB}M. Berciu and R. N. Bhatt, Phys.\ Rev.\ Lett.\ \textbf{87},
107203 (2001); \eprint{cond-mat/0111045}.


\bibitem{comment}C. Timm, F. Sch\"afer, and F. von Oppen, submitted to
Phys.\ Rev.\ Lett.\ (comment), \eprint{cond-mat/0111504};
M. Berciu and R. N. Bhatt, submitted to Phys.\ Rev.\ Lett.\ (reply),
\eprint{cond-mat/0112165}.


\bibitem{SE}B. I. Shklovskii and A. L. Efros, \textit{Electronic Properties
of Doped Semiconductors}, Solid-State Sciences \textbf{45} (Springer,
Berlin, 1984).

\bibitem{Keldysh}L. V. Keldysh and G. P. Proshko, Sov.\ Phys.---Solid State
\textbf{6}, 1093 (1964--1965).

\bibitem{rem.rscr}The electronic screening length $r_{\mathrm{scr}}$
decreases for reduced disorder potential. However, $r_{\mathrm{ion}}$
is always the smaller, and thus relevant, length scale.


\bibitem{Schott}G. M. Schott, W. Faschinger, and L. W. Molenkamp, Appl.\
Phys.\ Lett.\ \textbf{79}, 1807 (2001).

\bibitem{Shioda}R. Shioda \textit{et al.}, Phys.\ Rev.\
B \textbf{58}, 1100 (1998).

\bibitem{Beschoten}B. Beschoten \textit{et al.}, Phys.\ Rev.\ Lett.\
\textbf{83}, 3073 (1999).

\bibitem{rem.KL}For material-specific calculations the detailed band
structure should be taken into account \protect\cite{Abolfath,Dietl}. This
should not change the qualitative results but is known to increase the
mean-field $T_c$ \protect\cite{Dietl}.

\bibitem{rem.number}We have checked that the results do not change
significantly if the number of plane waves is increased.


\bibitem{Matsukura}F. Matsukura \textit{et al.}, Phys.\ Rev.\ B
\textbf{57}, R2037 (1998).

\bibitem{Okaba}J. Okabayashi \textit{et al.}, Phys.\ Rev.\ B \textbf{58},
R4211 (1998).

\bibitem{Bhatta}A. K. Bhattacharjee and C. Benoit \`a la Guillaume, Solid
State Commun.\ \textbf{113}, 17 (2000).

\bibitem{Yang}S.-R. E. Yang and A. H. MacDonald (submitted to Phys.\ Rev.\
Lett.), \eprint{cond-mat/0202021}.

\end{thebibliography}
\end{document}